\documentclass{iopart}
\usepackage{graphicx}
\usepackage{subfigure}
\usepackage{bm}
\usepackage{cite}



\begin{document}
\article[Rapidity-dependent chemical potentials]{Strangeness in Quark Matter, Levo\v ca, Slovakia, June 2007}{Rapidity-dependent chemical potentials \\in a statistical 
approach\footnote{Research supported by the Polish Ministry of Education and Science, grants N202~034~32/0918 and 2~P03B~02828.}}
\author{Wojciech Broniowski}
\address{The H. Niewodnicza\'nski Institute of Nuclear Physics, \\Polish Academy of Sciences, PL-31342 Krak\'ow, Poland}
\address{Institute of Physics, \'Swi\c{e}tokrzyska Academy, PL-25406~Kielce, Poland} 
\ead{Wojciech.Broniowski@ifj.edu.pl}

\author{Bart\l{}omiej Biedro\'n}
\address{AGH University of Science and Technology, PL-30059 Krak\'ow, Poland}
\ead{rockhouse@dione.ifj.edu.pl}

\begin{abstract}
We present a single-freeze-out model with thermal and geometric parameters dependent on the
position within the fireball and use it to describe the rapidity 
and transverse-momentum spectra of pions, kaons, protons, and antiprotons measured at RHIC 
at $\sqrt{s_{NN}}=200~{\rm GeV}$ by BRAHMS. {\tt THERMINATOR} is used to perform the
necessary simulation, which includes all resonance decays. The result of the fit to 
the data is the expected growth of the baryon and strange
chemical potentials with the spatial rapidity $\alpha_\parallel$. The value of the baryon chemical potential 
at $\alpha_\parallel \sim 3$ is about $200~{\rm MeV}$, 
{\em i.e.} lies in the range of the highest SPS energies. 
The chosen geometry of the fireball
has a decreasing transverse size as the magnitude of $\alpha_\parallel$ 
is increased, which also corresponds to decreasing transverse flow. 
The strange chemical potential obtained from the fit to the $K^+/K^-$ ratio is such that
the local strangeness density in the fireball is compatible with zero.
The resulting rapidity spectra of net protons are described qualitatively within the statistical approach.  
As a result of our study, the knowledge of the ``topography'' of the fireball is acquired, 
allowing for other analyses and predictions.
\end{abstract}


\submitto{\JPG}

\date{31 August 2007}

\maketitle

More details of the presented material can be found in Ref.~\cite{BBWB}.

An important goal of relativistic heavy-ion physics is to find the ``topography'' of the system at freeze-out, {\em i.e.} 
to determine the dependence of thermal parameters on the location, in particular on the 
spatial rapidity $\alpha_\parallel=\frac{1}{2} \log{(t-z)/(t+z)}$. 
Numerous studies of the global particle ratios \cite{Koch:1985hk,Cleymans:1992zc,Sollfrank:1993wn,Braun-Munzinger:1994xr,%
Gazdzicki:1998vd,Yen:1998pa,Cleymans:1998fq,%
Becattini:2000jw,Rafelski:2000by,Broniowski:2001we,Broniowski:2002nf,Becattini:2003wp,Braun-Munzinger:2003zd,Torrieri:2003nh,%
Cleymans:2004pp,Rafelski:2004dp,Becattini:2005pt}
can be divided into two categories: the $4\pi$ studies at low
energies (SIS, AGS) and the analyses at mid-rapidity for approximately
boost-invariant systems at high energies (RHIC). 
In the latter case the yields $dN/dy$ 
vary only by a few percent in the window $|y| < 1$. 
When the system is not boost-invariant the task is more involved because particles
detected at a given pseudorapidity may originate
from different locations in the fireball. Thermal conditions
and flow change from place to place, which should be
taken into account. In addition, the resonance decays are more complicated to treat. 
As a result, a full-fledged simulation is necessary. 
We use {\tt THERMINATOR} 
\cite{Kisiel:2005hn} with a suitably extended
single-freeze-out model of Ref.~\cite{Broniowski:2001we,Broniowski:2001uk,Broniowski:2002nf}.
We recall the approach reproduces in an efficient way the transverse-momentum spectra,
including strange particles \cite{Broniowski:2001uk}, the production of resonances \cite{Broniowski:2003ax},
the charge balance functions \cite{Bozek:2003qi}, the elliptic flow 
\cite{Broniowski:2002wp}, the HBT radii \cite{Kisiel:2006is}, and the transverse
energy \cite{Prorok:2004wi,Prorok:2005uv}. To fix the model parameters we use the data from the BRAHMS collaboration at 
$\sqrt{s_{NN}}=200$~GeV \cite{Bearden:2003fw,Bearden:2004yx,Murray,Staszel}.

Our extension to boost-non-invariant
systems consists of two elements: the choice of the freeze-out hypersurface and the collective flow, as well as 
incorporation of the dependence of thermal parameters on the location. 
We use the freeze-out hypersurface parameterized as
\begin{eqnarray}\!\!\!\!\!\!\!\!
x^\mu = \tau \left( 
\mathrm{cosh} \alpha_\perp \mathrm{cosh } \alpha_\parallel,  
\mathrm{sinh} \alpha_\perp \cos\phi, \mathrm{sinh} \alpha_\perp \sin\phi, 
\mathrm{cosh} \alpha_\perp \mathrm{sinh } \alpha_\parallel \right).  \label{x}
\end{eqnarray}
The parameter $\tau$ is the proper time at freeze-out, and $\alpha_\perp$ is related to the transverse radius,
$\rho = \sqrt{x^2+y^2} = \tau \mathrm{sinh} \alpha_\perp$.
The four-velocity is chosen to follow the Hubble flow
$u^\mu = x^\mu/\tau.  \label{u}$
In the boost-invariant model $\rho$ was limited by the
space-independent parameter $\rho_{\mathrm{max}}$, or $0 \le \alpha_\perp
\le \alpha_\perp^{\mathrm{max}}$. Now we take 
\begin{eqnarray}
0 \le \alpha_\perp \le \alpha_\perp^{\mathrm{max}}(\alpha_\parallel) \equiv
\alpha_\perp^{\mathrm{max}}(0) \exp \left ( -\frac{\alpha_\parallel^2}{2
\Delta^2}\right ).  \label{eq:Delta}
\end{eqnarray}
The interpretation of Eq.~(\ref{eq:Delta}) is following: as we depart from the center,
increasing $|\alpha_\parallel|$, we simultaneously reduce $\alpha_\perp$, or 
$\rho_{\mathrm{max}}$. The rate of this reduction is controlled by a new
model parameter, $\Delta$. In other words, the fireball gets thinner and thinner as we depart from 
the central rapidity region.
The geometry and expansion of the fireball are described by three
parameters: $\tau$, controlling the overall abundance of particles, $\rho_{\mathrm{max}}^{(0)}$, influencing the slope of the
$p_T$-spectra, and $\Delta$, controlling the fall-off of the rapidity spectra.
With the standard parameterization of the particle four-momentum in terms of
rapidity and the transverse mass, 
$p^\mu = \left(m_\perp \hbox{cosh} y, p_\perp \cos\varphi, p_\perp
\sin\varphi, m_\perp \hbox{sinh} y \right),$
we find
\begin{eqnarray}
p \cdot u &=& m_\perp \hbox{cosh}(\alpha_\perp) \hbox{cosh}(\alpha_%
\parallel-y) - p_\perp \hbox{sinh}(\alpha_\perp) \cos(\phi-\varphi), \nonumber \\ 
d^3\Sigma \cdot p &=& \tau^3 d\alpha_\parallel d\phi \,\mathrm{sinh}\alpha_\perp \,
d\alpha_\perp \, p \cdot u  \label{sigmapbinv}
\end{eqnarray}
where $d^3\Sigma^\mu$ is the volume element of the hypersurface. These expressions are used the Cooper-Frye \cite{Cooper:1974mv} formula
for the momentum density of a given species
of {\em primordial} particles: 
\begin{eqnarray}
\frac{d^2N}{ 2\pi p_T dp_T \,dy} &=& \tau^3 \!\! \int_{-\infty}^\infty
\!\!\!\!\!\! d\alpha_\parallel \int_0^{\alpha_\perp^{\mathrm{max}%
}(\alpha_\parallel)}\!\!\!\!\!\!\!\!\!\!\!\!\! d\alpha_\perp \int_0^{2\pi}
\!\!\!\!\!\! d\phi  p \cdot u f\left ( \beta p \cdot u - \beta
\mu(\alpha_\parallel)\right ),  \label{master}
\end{eqnarray}
where $p \cdot u$ is taken at $\varphi = 0$ (azimuthal symmetry), $f$
denotes the Bose-Einstein or Fermi-Dirac statistical distribution function, $\beta=1/T$, and 
\mbox{$\mu(\alpha_\parallel) =B \mu_B(\alpha_\parallel) + S \mu_S(\alpha_\parallel)+I_3 \mu_{I_3}(\alpha_\parallel)$},
with $B$, $S$, and $I_3$ denoting the baryon number, strangeness, and the
third component of isospin of the given particle. Thus we admit the dependence of
chemical potentials on $\alpha_\parallel$. 
In general, the temperature $T$ also depends on $\alpha_\parallel$. In this work, however, we
apply the model for not too large values of rapidity, $|y| \le 3.3$, where 
the obtained values for $\mu_B$ are less than $\sim 250~%
\mathrm{MeV}$. The universal freeze-out curve \cite{Cleymans:1998fq} gives  a practically 
constant value of $T$ in this range, thus we fix $T=165~\mathrm{MeV}$ everywhere.

\begin{figure}[tb]
\begin{center}
\subfigure{\includegraphics[width=.4\textwidth]{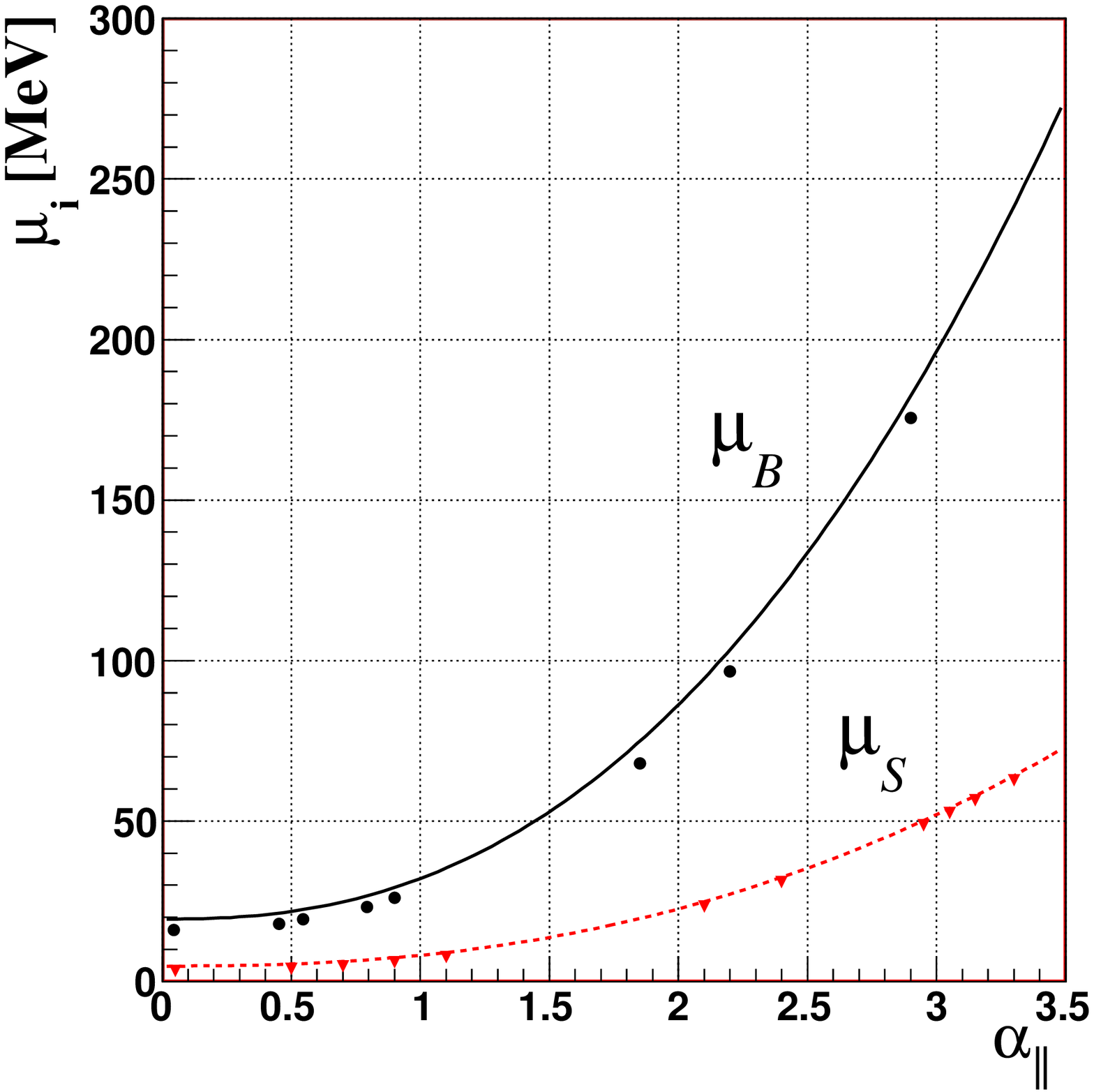}}
\subfigure{\includegraphics[width=.43\textwidth]{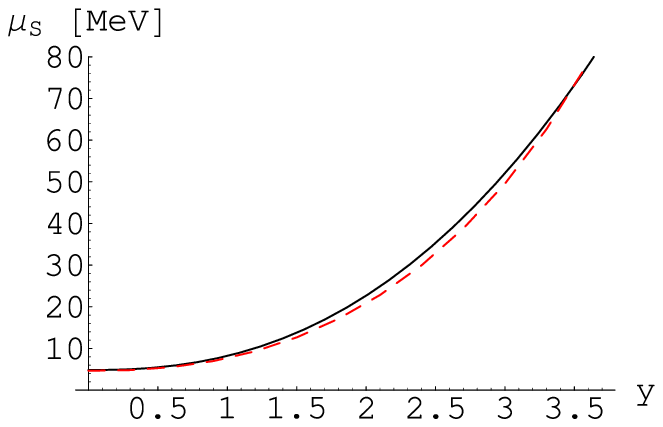}}
\end{center}
\vspace{-5mm}
\caption{
Left: the model baryon and strange chemical potentials 
plotted as functions of $\alpha_\parallel$. 
Parameters of Eq.~(\ref{ff}) are obtained from the fit to the 
BRAHMS data \protect\cite{Bearden:2003fw,Bearden:2004yx}. The points represent 
a naive calculation based of Eq.~(\ref{start}). Right: 
Comparison of $\mu_S$ obtained from the fit to the data (solid line) and 
from $\mu_B$ and the condition of zero local strangeness density (dashed line). \label{fig:sc}
\label{fig:muy}}
\end{figure}
   
The functional form of the chemical potentials is parameterized as follows: 
\begin{equation}
\mu _{i}(\alpha _{\parallel })=\mu _{i}(0) \left [1+A_{i}\alpha _{\parallel}^{2.4} \right],\;\;\; i=B,S,I_{3}. \label{ff}
\end{equation}
Practice shows that to a good approximation the statistical distributions may be replaced by
the Boltzmann factors. Then the integrand of Eq.~(\ref%
{master}) contains the term $\exp[-\beta m_\perp \mathrm{cosh}%
(\alpha_\perp) \mathrm{sinh}(\alpha_\parallel - y) + \beta
\mu(\alpha_\parallel)]$. The relevant integration range in $%
\alpha_\parallel$ is sharply peaked around $\alpha_\parallel \simeq y$, thus the chemical
potentials are taken approximately at $\mu_i(\alpha_\parallel) \simeq \mu_i(y)$ and the factors $\exp[\beta \mu(y)]$ 
can be pulled out from the integration. As a result, the following  
{\em approximate} relations for the ratios of yields hold: 
\begin{eqnarray}
\frac{p}{\bar p} \simeq \exp (2\beta\mu_B), \;\;\frac{K^+}{K^-} \simeq \exp (2\beta\mu_S), \;\;
\frac{\pi^+}{\pi^-} \simeq \exp (2\beta\mu_{I_3}).  \label{start}
\end{eqnarray}
Upon inversion $\mu_B(y)\simeq \frac{1}{2} T \log (p/{\bar p})$, {\em etc.}
With the help of these approximate equalities we set the starting values of the
parameters $\mu_i(0)$ and $A_i$, which are then iterated by {\tt THERMINATOR} to fit the data. 
The $\Delta$ parameter is fixed with the pion rapidity spectra 
$dN_{\pi^\pm}/dy$, with the optimum value of $\Delta=3.33$. 
The other geometric parameters are taken from earlier fits to mid-rapidity $p_T$-spectra: $\tau = 9.74$~fm,
$\rho_{\rm max}(z = 0) = 7.74$~fm.

The result of our optimization is shown in 
Fig.~\ref{fig:muy}. The optimum parameters are:
\begin{eqnarray}
&& \!\!\!\!\!\!\!\! \mu _{B}(0)=19~{\rm MeV},
\;\mu _{S}(0)=4.8~{\rm MeV},\;\mu _{I_3}(0)=-1~{\rm MeV},  \nonumber \\
&& \!\!\! A_{B}=0.65,\;A_{S}=0.70,\;A_{I_3}=0. \label{numpar}
\end{eqnarray}%
\begin{figure}[tb]
\begin{center}
\vspace{-1mm}
\subfigure{\includegraphics[width=.4\textwidth]{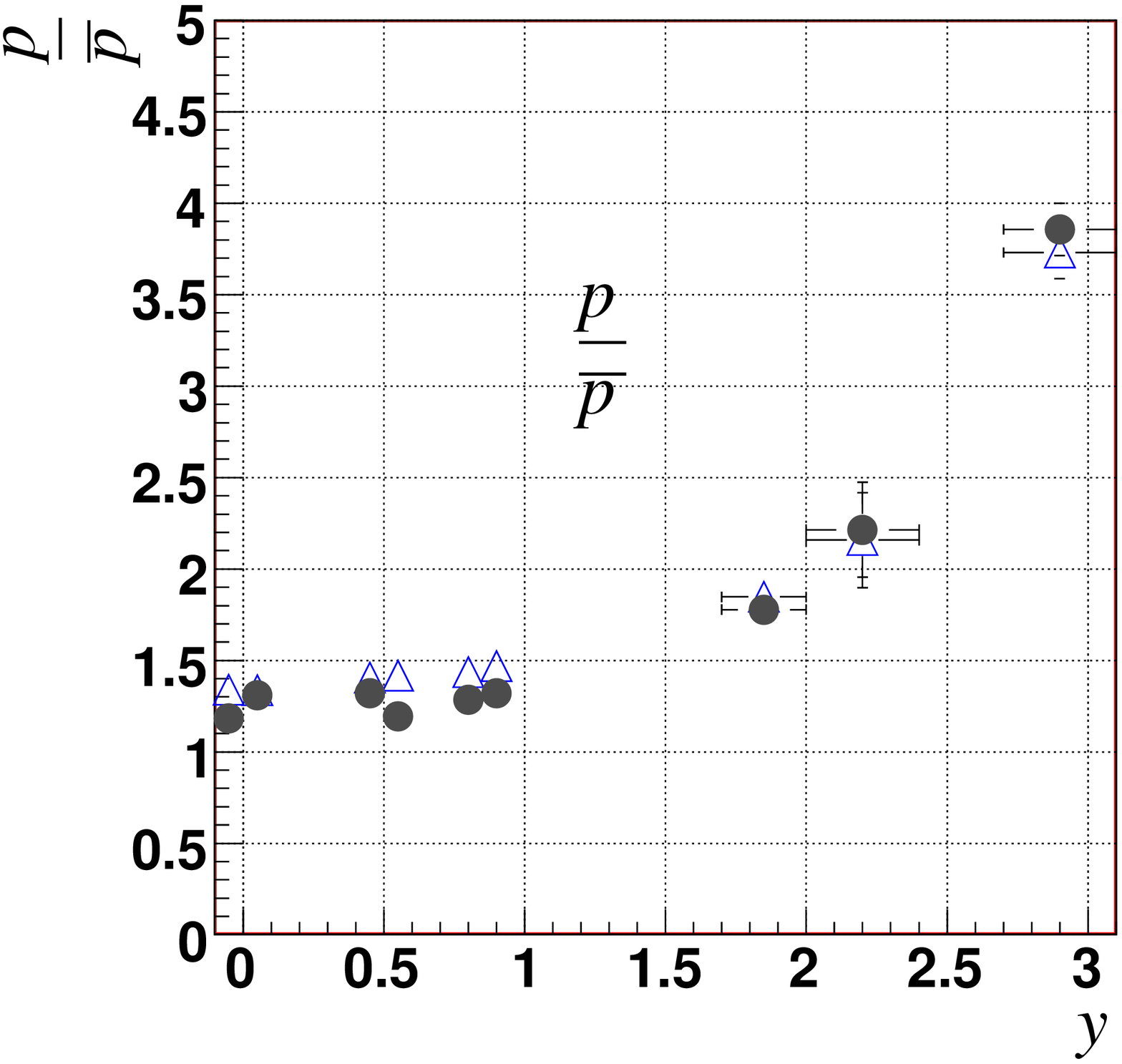}}
\subfigure{\includegraphics[width=.4\textwidth]{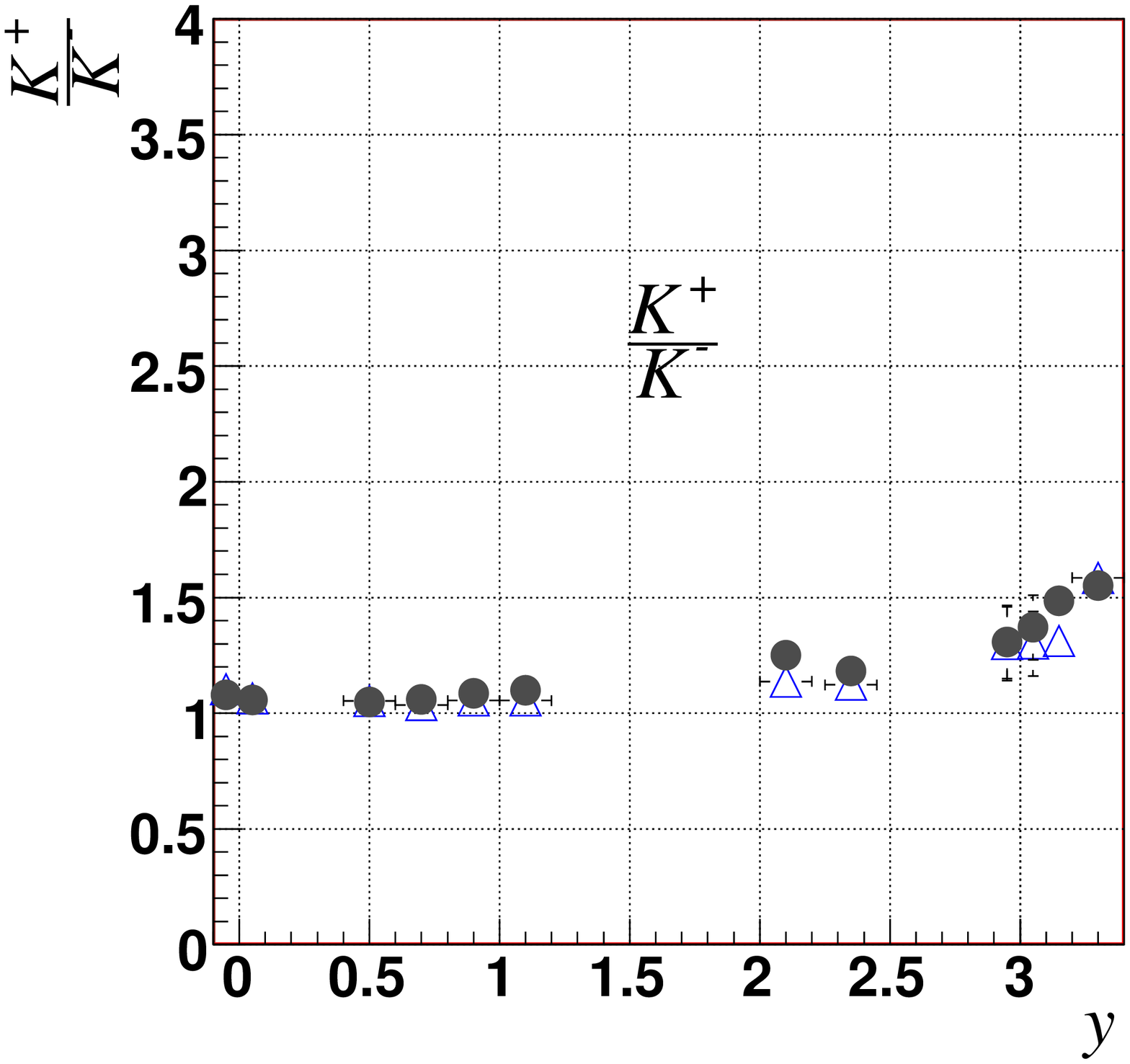}}\\
\end{center}
\vspace{-9mm}
\caption{Dependence of particle ratios on rapidity: left -- $p/%
{\bar p}$, right -- $K^+/K^-$.
The open triangles are the Brahms data \protect\cite%
{Bearden:2003fw,Bearden:2004yx}, while the filled circles show the result of the
model simulation. The data for $p$ and $\bar p$ 
are not corrected for the feed-down from weak-decays \cite{Bearden:2003fw}.
\label{fig:ratiosy}}
\end{figure}
We observe the expected behavior for the baryon chemical potential, which
increases with $|\alpha _{\parallel }|$. The value at the origin is 19~MeV, 
somewhat lower than the earlier mid-rapidity fits made
in boost-invariant models in 
Refs.~\cite{Broniowski:2001we,Braun-Munzinger:2001ip}, yielding 26~MeV. 
The lower value in our case is well understood. The previous mid-rapidity fits 
include the data in the 
range $|y| \le 1$. This range collects the particles emitted from the fireball at 
$|\alpha_\parallel| \le 2$, hence the value of $\mu_B$ in the 
previous mid-rapidity fits is an average of 
our $\mu_B(\alpha_\parallel)$ over the range, approximately, $|\alpha_\parallel| \le 2$, 
with some weight proportional to the particle abundance. 
A similar effect occurs for $\mu_S$.
We do not incorporate corrections for the feed-down from weak decays ({\em i.e.} 
all the decays are 
included), since this is the policy of Ref.~\cite{Bearden:2003fw} for the
treatment of $p$ and $\bar p$.

We note that at $\alpha _{\parallel }=3$
the value of $\mu_B$ is 200~MeV. This value is comparable to the 
highest-energy SPS fit ($\sqrt{s_{NN}}=17~\mathrm{GeV}$), where $\mu
_{B}\simeq 230~\mathrm{MeV}$. The behavior of the strange chemical
potential is qualitatively similar. It also increases with $|\alpha
_{\parallel }|$, growing from 5~MeV at the origin to 50~MeV at $\alpha _{\parallel }=3$.
The ratio $\mu_B(\alpha_\parallel)/\mu_S(\alpha_\parallel)$ 
is very close to a constant $\simeq 4-3.5$.
The dots in the left panel of Fig.~\ref{fig:muy} show the result of 
the naive calculation of Eq.~(\ref{start}). We note that these points are 
close to the result of the full-fledged fit of our model. 
This is of practical use, since the application of Eq.~(\ref{start}) 
involves no effort (as done {\em e.g.} in Ref.~\cite{Cl0,Cl,BC}), while the full 
model simulation incorporating resonance decays, flow, {\em etc.}, is costly.

There is another interesting point. In thermal models one may obtain the 
local value of the 
strange chemical potential, $\mu_S$, at a given $\mu_B$ with the condition of 
the vanishing strangeness density, $\rho_S=0$. 
The result is shown in Fig.~\ref{fig:sc}, where we compare the strange chemical 
potential obtained from the fit to the data (solid line) and 
from the condition of zero local strangeness density at 
a given $\mu_B(\alpha_\parallel)$ (dashed line). 
The two curves
turn out to be virtually the same. This shows that the 
net strangeness density in our fireball is, within uncertainties of parameters, 
{\em compatible with 0}. This is not obvious from the outset, 
as the condition of zero strangeness density is not assumed in our fitting procedure.
Although this feature is natural in particle 
production mechanisms, in principle only the total strangeness,
integrated over the whole fireball, must be initially zero. 
Variation of the strangeness density with $\alpha_\parallel$ is admissible, 
but turns out not to occur. 
 
\begin{figure}[tb]
\begin{center}
\subfigure{\includegraphics[width=.4\textwidth]{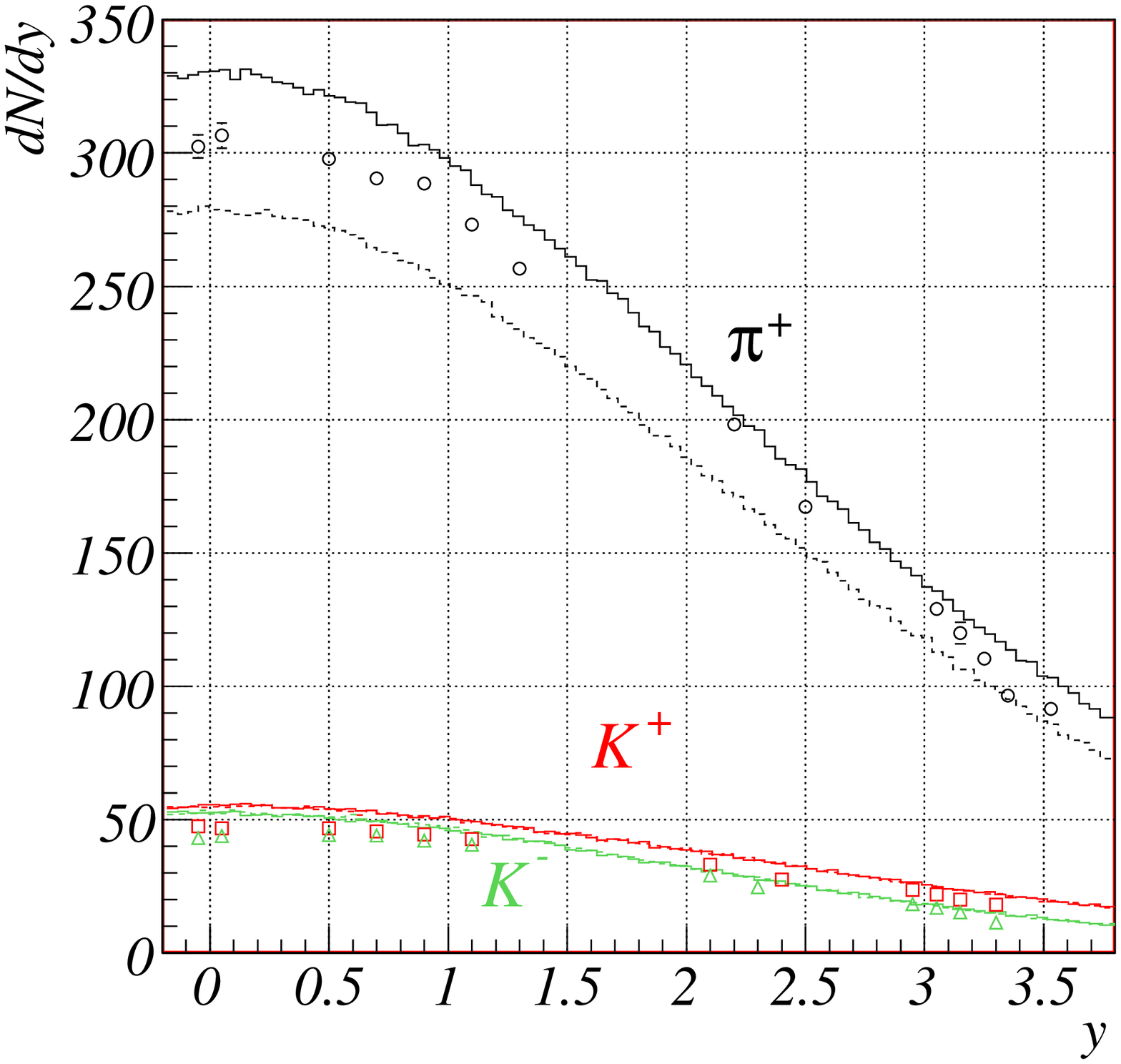}}
\subfigure{\includegraphics[width=.4\textwidth]{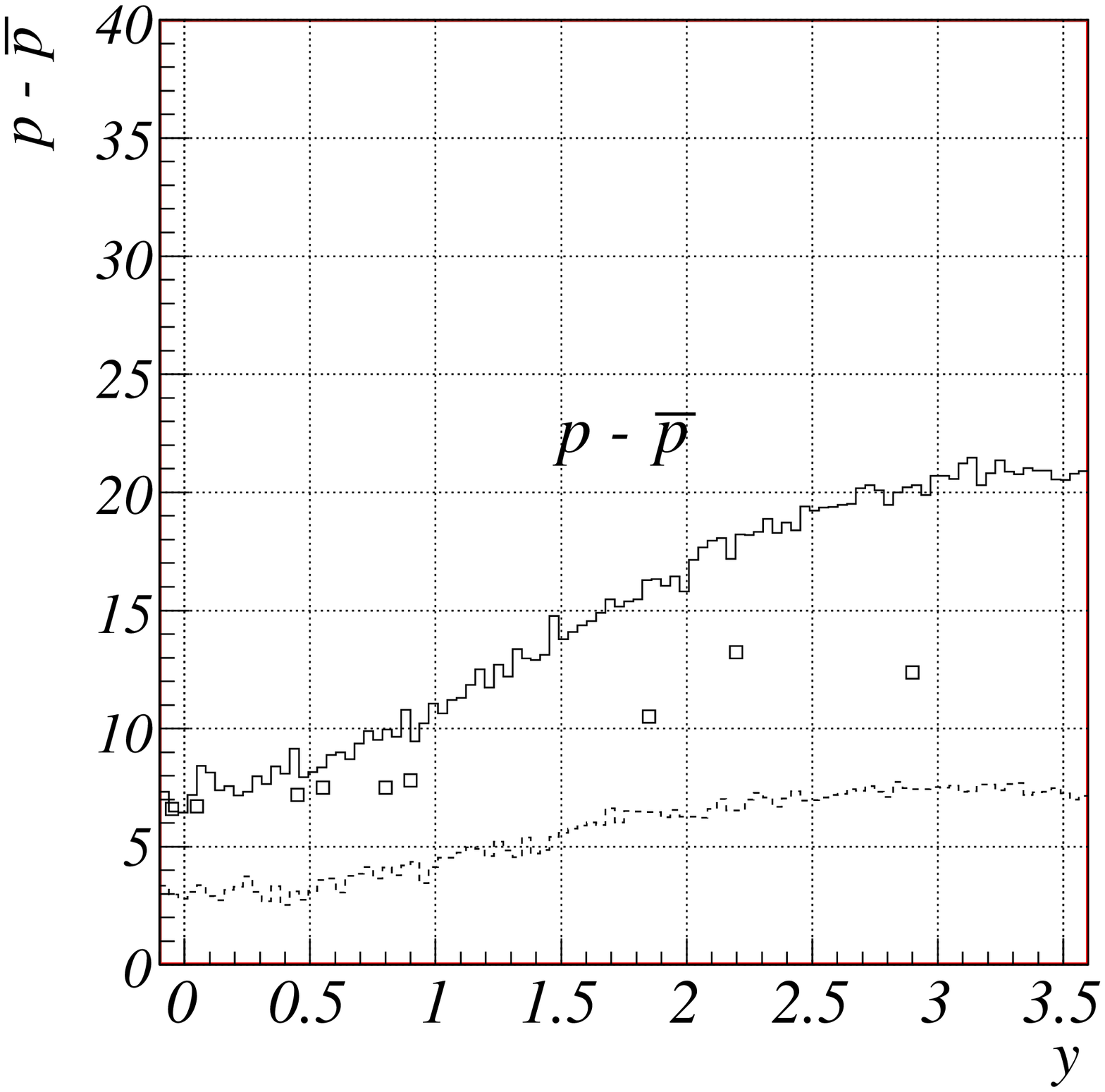}}
\end{center}
\vspace{-6mm}
\caption{Left: 
Left: Rapidity spectra of $\pi^+$, $K^+$, and $K^-$. The data come
from  Refs.~\cite{Bearden:2003fw,Bearden:2004yx} 
(circles -- $\pi^+$, squares -- $K^+$, triangles -- $K^-$),
while the histogram lines show the model results. 
For $\pi^+$ the solid (dashed) line corresponds to the 
full feeding (no feeding) from the weak hyperon decays. 
The experimental pion yields are corrected 
for weak decays \cite{Bearden:2004yx}.
Right: spectrum of net protons, $p- \bar p$.
The data points come 
from Refs.~\cite{Bearden:2003fw,Bearden:2004yx},
while the solid (dashed) histogram lines show the result of the model simulation 
with full feeding (no feeding) from the weak hyperon decays. 
Data points should be compared to the model with full feeding (solid lines).
\label{fig:bry}\label{fig:bry2}}
\end{figure}

Figure \ref{fig:ratiosy} shows the quality of our fit of the 
parameters of the chemical potentials, 
Eq.(\ref{numpar}), for the ratios  $p/{\bar p}$ and $K^+/K^-$. 
In Fig.~\ref{fig:bry} we compare the obtained 
rapidity spectra of $\pi^+$, $K^+$, and $K^-$ to the experiment. 
The experimental yields for the pions
are corrected for the feed-down from the weak decays as described in 
Ref.~\cite{Bearden:2004yx}. For that reason
for the case of 
$\pi^+$ we give the model predictions with the full feeding from the weak 
decays (solid line)
and with no feeding at all (dashed line). We note a
reasonable agreement, with the data 
falling between the two extreme cases. 
The spectra of kaons are also reproduced reasonably well.    
We see from the right panel of Fig.~\ref{fig:bry2} that 
the qualitative growing of the net-proton spectrum with $y$ is reproduced. 
Note that predictive power is left for this observable, as only the ratio $\bar p/p$ is used to fit 
$\mu_B$. The model curves with and without the weak-decays correction embrace the experimental points. 

Here are the highlights this talk:

\begin{enumerate}

\item Decreasing yields of particles with rapidity require that the system becomes colder, thinner, or more dilute as one 
departs from mid-rapidity. In our analysis for the highest RHIC energies we take it to be thinner, keeping the temperature constant.  

\item The naive extraction of the baryon and strange chemical 
potentials from ratios of $p/\bar p$ and $K^+/K^-$ according to Eq.~(\ref{start}) works 
surprisingly well, as shown in the comparison to the 
full calculation in Fig.~\ref{fig:muy}. This is because at $\sqrt{s_{NN}}=200$~GeV the dependence of yields 
on $\alpha_\parallel$ does not depart much from linearity within one unit of rapidity, which is the window contributing to the 
$\alpha_\parallel$ integration in Eq.~(\ref{master}). 
In addition, the feeding contributions from resonance decays to both particles in the ratio are approximately proportional 
to each other. At lower energies (SPS) the naive extraction should not work.

\item The baryon and strange chemical potentials grow 
with $\alpha_\parallel$, reaching at $y \sim 3$ values 
close to those of the highest SPS energies of $\sqrt{s_{NN}}=17~{\rm GeV}$.
This agrees with the conclusions of Roehrich \cite{Roehrich}. 

\item At mid-rapidity the values of the chemical potentials are even lower
than derived from the previous thermal fits to the data for $|y|\le 1$, with 
our values taking $\mu_B(0)=19~{\rm MeV}$ and $\mu_S(0)=5~{\rm MeV}$. 
The reason for this effect 
is that the particle with $|y|\le 1$ originate from a region $|\alpha_\parallel|\le
2$, and on the average the effective values of chemical potentials are larger 
compared to the values at the very origin
(cf. Fig.~\ref{fig:muy}).

\item The local strangeness density of the fireball is compatible with zero at all 
values of $\alpha_\parallel$. Although this feature is natural in particle 
production mechanisms, here it has been obtained independently just from 
fitting the chemical potentials to the data.

\item The ratio of the baryon to strange chemical potentials varies very weakly with 
rapidity, ranging from $\sim 4$ at midrapidity to $\sim 3.5$ at larger rapidities.

\item The $d^2 N/(2\pi p_\perp dp_\perp dy)$ spectra of pions 
and kaons are also well reproduced \cite{BBWB}, complying to our hypothesis for the shape of the fireball 
in the longitudinal direction.

\item  The feature of the
increasing yield of the net protons, $p-\bar p$, with rapidity is obtained naturally, 
explaining qualitatively the shape of the rapidity dependence 
on purely statistical grounds. 

\end{enumerate} 

\vfill


\begin{thebibliography}{10}
\expandafter\ifx\csname url\endcsname\relax
  \def\url#1{\texttt{#1}}\fi
\expandafter\ifx\csname urlprefix\endcsname\relax\def\urlprefix{URL }\fi

\bibitem{BBWB}
B.~Biedro\'n, W.~Broniowski, Phys. Rev. C75 (2007) 054905.

\bibitem{Koch:1985hk}
P.~Koch, J.~Rafelski, South Afr. J. Phys. 9 (1986) 8.

\bibitem{Cleymans:1992zc}
J.~Cleymans, H.~Satz, Z. Phys. C57 (1993) 135.

\bibitem{Sollfrank:1993wn}
J.~Sollfrank, M.~Gazdzicki, U.~W. Heinz, J.~Rafelski, Z. Phys. C61 (1994) 659.

\bibitem{Braun-Munzinger:1994xr}
P.~Braun-Munzinger, J.~Stachel, J.~P. Wessels, N.~Xu, Phys. Lett. B344 (1995) 43.

\bibitem{Gazdzicki:1998vd}
M.~Gazdzicki, M.~I. Gorenstein, Acta Phys. Polon. B30 (1999) 2705.

\bibitem{Yen:1998pa}
G.~D. Yen, M.~I. Gorenstein, Phys. Rev. C59 (1999) 2788.

\bibitem{Cleymans:1998fq}
J.~Cleymans, K.~Redlich, Phys. Rev. Lett. 81 (1998) 5284.

\bibitem{Becattini:2000jw}
F.~Becattini, J.~Cleymans, A.~Keranen, E.~Suhonen, K.~Redlich, Phys. Rev. C64 (2001)
  024901.

\bibitem{Rafelski:2000by}
J.~Rafelski, J.~Letessier, Phys. Rev. Lett. 85 (2000) 4695--4698.

\bibitem{Broniowski:2001we}
W.~Broniowski, W.~Florkowski, Phys. Rev. Lett. 87 (2001) 272302.

\bibitem{Broniowski:2002nf}
W.~Broniowski, A.~Baran, W.~Florkowski, Acta Phys.
  Polon. B33 (2002) 4235--4258.

\bibitem{Becattini:2003wp}
F.~Becattini, M.~Gazdzicki, A.~Keranen, J.~Manninen, R.~Stock,
  Phys. Rev. C69 (2004) 024905.

\bibitem{Braun-Munzinger:2003zd}
P.~Braun-Munzinger, K.~Redlich, J.~Stachel, in Quark Gluon
Plasma 3, ed. R.~C.~Hwa and X.~N.~Wang (World Scientific,
Singapore, 2004).

\bibitem{Torrieri:2003nh}
G.~Torrieri, J.~Rafelski, J. Phys. G30 (2004) S557.

\bibitem{Cleymans:2004pp}
J.~Cleymans, B.~Kampfer, M.~Kaneta, S.~Wheaton, N.~Xu, Phys. Rev. C71 (2005) 054901.

\bibitem{Rafelski:2004dp}
J.~Rafelski, J.~Letessier, G.~Torrieri, Phys. Rev. C72 (2005) 024905.

\bibitem{Becattini:2005pt}
F.~Becattini, L.~Ferroni, J. Phys. G31 (2005) S1091.

\bibitem{Kisiel:2005hn}
A.~Kisiel, T.~Taluc, W.~Broniowski, W.~Florkowski, Comput.
Phys. Commun. 174 (2006) 669.

\bibitem{Broniowski:2001uk}
W.~Broniowski, W.~Florkowski, Phys. Rev. C65 (2002) 064905.

\bibitem{Broniowski:2003ax}
W.~Broniowski, W.~Florkowski, B.~Hiller, Phys. Rev. C68 (2003) 034911.

\bibitem{Bozek:2003qi}
P.~Bo\.zek, W.~Broniowski, W.~Florkowski, Acta Phys. Hung. A22 (2005) 149.

\bibitem{Broniowski:2002wp}
W.~Broniowski, A.~Baran, W.~Florkowski, AIP Conf. Proc. 660 (2003) 185.

\bibitem{Kisiel:2006is}
A.~Kisiel, W.~Florkowski, W.~Broniowski, Phys. Rev. C73 (2006) 064902.

\bibitem{Prorok:2004wi}
D.~Prorok, Eur. Phys. J. A26 (2005) 277.

\bibitem{Prorok:2005uv}
D.~Prorok, Phys. Rev. C73 (2006) 064901.

\bibitem{Bearden:2003fw}
I.~G. Bearden, et~al., Phys. Rev. Lett. 90 (2003) 102301.

\bibitem{Bearden:2004yx}
I.~G. Bearden, et~al., Phys. Rev. Lett. 94 (2005) 162301.

\bibitem{Murray} M. Murray, these proceedings.

\bibitem{Staszel} P. Staszel, these proceedings.

\bibitem{Cooper:1974mv}
F.~Cooper, G.~Frye, Phys. Rev. D10 (1974) 186.

\bibitem{Braun-Munzinger:2001ip}
P.~Braun-Munzinger, D.~Magestro, K.~Redlich, J.~Stachel, Phys. Lett. B518 (2001) 41.

\bibitem{Cl0}
J.~Cleymans, these proceedings.

\bibitem{Cl}
J.~Cleymans, these proceedings.

\bibitem{BC} F. Becattini, J. Cleymans, J. Phys. G34 (2007) S959. 

\bibitem{Roehrich}
D.~Roehrich, talk at Critical Point and Onset of Deconfinement
(Florence, 3–6 July, 2006).


\end{thebibliography}

\end{document}